\documentclass[reprint,amsmath,amssymb,aps,pra]{revtex4-1}

\usepackage{hyperref}
\usepackage{color}
\usepackage{graphicx}
\usepackage{dcolumn}
\usepackage{bm}
\usepackage{soul}
\usepackage{amsmath}

\newcommand{\ket}[1]{| #1 \rangle}
\newcommand{\bra}[1]{\langle #1 |}

\begin{document}

\title{Phase shift in atom interferometers: corrections for non-quadratic potentials and finite-duration laser pulses}

\author{A. Bertoldi} 
\affiliation{LP2N, Laboratoire Photonique, Num\'erique et
  Nanosciences,\\ Universit\'e Bordeaux-IOGS-CNRS:UMR 5298, F-33400 Talence, France.}

\author{F. Minardi} 
\email{francesco.minardi@ino.cnr.it}
\affiliation{Istituto Nazionale di Ottica, CNR-INO, 50019 Sesto Fiorentino, Italy}
\affiliation{LENS and Dipartimento di Fisica e Astronomia, Universit\`{a} di Firenze, 50019 Sesto Fiorentino, Italy}
\affiliation{Dipartimento di Fisica e Astronomia, Universit\`{a} di Bologna, 40127 Bologna, Italy}

\author{M. Prevedelli}
\affiliation{Dipartimento di Fisica e Astronomia, Universit\`{a} di Bologna, 40127 Bologna, Italy}

\date{\today}

\begin{abstract}
  We derive an expression for the phase shift of an atom
  interferometer in a gravitational field taking into account both the
  finite duration of the light pulses and the effect of a small
  perturbing potential added to a stronger uniform gravitational
  field, extending the well-known results for rectangular pulses and
  at most quadratic potentials. These refinements are necessary for a
  correct analysis of present day high resolution interferometers.
\end{abstract}


\maketitle


\section{\label{sec:intro}Introduction}

Atom Interferometry (AI) rests upon the coherent manipulation of
matter waves \cite{Borde_1989}. The increasing ability to control
individual quantum systems and their evolution makes it feasible to
observe quantum interference over trajectories with very large
separation in momentum \cite{Chiow_2011,PlotkinSwing_2018} and space
\cite{Kovachy_2015}. The resulting high sensitivity and the exquisite
control of systematic effects are at the basis of the growing number
of applications in AI, ranging from tests of general relativity
\cite{Dimopoulos_2007,Canuel2018}, measurement of fundamental
constants \cite{Rosi_2014,Parker_2018}, search for new physics
\cite{Hamilton_2015,Jaffe_2017}, to more applied contexts like
inertial navigation \cite{Cheiney_2018}.

The improving experimental performances of AI require a refinement of
the modeling for the phase shift calculation. Two main formulations
have been developed to obtain the interferometric phase shift in the
case of two-path configurations, with three or more light pulses: a
path integral approach
\cite{Kasevich_1991,Storey_1994,Antoine_2003b,Bongs_2006,Cadoret_2016}, and a
density matrix equation in the Wigner representation
\cite{Dubetsky_2006,Dubetsky_2018}. Several effects have been
investigated especially in the first formulation, such as the finite
speed of light \cite{Cheng_2015} or the wavefront aberration of the
light beams \cite{Schkolnik_2015}. The calculation has been also
extended to the general relativistic case
\cite{Bord_2004,Dimopoulos_2008}.

We adopt here a formalism based on the Heisenberg representation to
describe the dynamics of a two level atom in an external potential
coherently manipulated with a pulsed laser beam \cite{Marzlin_1996};
this formulation provides the interferometric phase by adopting a
series of unitary transformations to write the evolution operator in
simple terms. First, we calculate the dependence of the
interferometric phase on the finite pulse duration, previously treated
in \cite{PetersPHD,Antoine_2006,Li_2015}: the result in
Eq. \ref{eq:phi2expanded} agrees with the existing literature, and is
valid for pulses of arbitrary shape. Our approach can be extended to
calculate the cumulated high order corrections imposed by multi--pulse
sequences adopted to increase the momentum separation of the
interfering trajectories \cite{Chiow_2011,PlotkinSwing_2018} or to
enhance the instrument sensitivity at a specific frequency
\cite{Graham_2016}. Second, we analyze the effect of more than
quadratic external potentials in atom interferometers, a problem for
which only a numerical solution has been proposed to date
\cite{Roura_2014}. Small terms beyond uniform gravity are treated with
perturbation theory, and the well known case of the quadratic
potential is used to validate our formulation. We demonstrate that the
so-called `sensitivity function' (SF) in AI
\cite{Cheinet_2008} gives the correct phase shift when the average
over the initial velocity distribution is considered, even if it
neglects a term of the Hamiltonian. We can also reinterpret the main
phase shift terms in the commonly adopted path integral description of
AI \cite{Storey_1994}. Evaluating the phase contribution of more than
quadratic terms of the gravitational potential is relevant to several
experiments where atoms are coherently manipulated close to the source masses
\cite{Hohensee_2012,Rosi_2014,Asenbaum_2017,Haslinger_2017,Jaffe_2017}.

The article is organized as follows: we describe our method based on
the Heisenberg picture in Sec. \ref{sec:meanpath}, where we consider
the frequency chirp required to maintain the manipulation laser on
resonance with the atoms, and implement the unitary transformation
that transfers the two interferometer trajectories on the classical
mean path. Sec. \ref{sec:quadPot} analyzes the well known case of AI
in a quadratic potential, and adopts another unitary transformation to
separate the effects on the interferometric output due to the free
evolution and to the pulses; the findings are compared with those
reported in the literature. In Sec. \ref{s:ppot} we consider the
effect of a more than quadratic external potential with a perturbative
theory, and generalize to arbitrary perturbative potentials the method
proposed in \cite{Roura_2017} to mitigate the contrast loss due to the
gravity-gradient.

\section{\label{sec:meanpath}Unitary transformation: mean path}

\begin{figure}[t!]
  \includegraphics[scale=1]{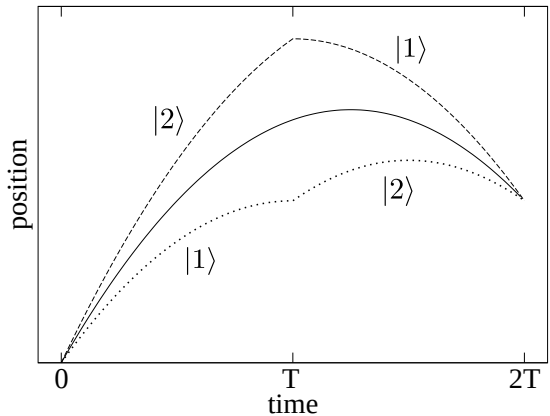}    
  \caption{\label{fig:paths}Classical trajectories in a Kasevich-Chu
    interferometer. The dashed and dotted lines represent the upper
    and lower interfering paths while the continuous line is the
    mean path, i.e. the trajectory of a particle with initial
    average momentum. At $t=T$ the $\pi$ pulse exchanges the internal
    states and momentum with respect to the mean path. For a linear
    potential $V(z)$ the three trajectories converge to the same point
    at $t=2T$; this condition is in general not valid for a nonlinear
    potential.}
\end{figure}

In order to focus on the essential features of the calculation, we
adopt a simplified two-level model in one dimension. Raman transitions
between two stable levels $\ket{1}$ and $\ket{2}$ are characterized by
a time-dependent Rabi frequency $\Omega(t)$, after adiabatic
elimination of the excited level. The atoms are initially prepared in
the internal state  $\ket{1}$ and their initial wavefunction is assumed to be a
Gaussian wavepacket in momentum. The atoms have been prepared with an
initial velocity selection pulse of length $\tau_s$, that fixes the
momentum distribution width.

The Hamiltonian describing the effective two-level atom interacting with the Raman laser beams is \cite{Marzlin_1996}
\begin{equation}
  H=\left[\frac{\hat{p}^2}{2m}+V(\hat{z})\right]I-\hbar \Omega(t) \cos \phi_L(\hat{z},t) \sigma_1+\frac{\hbar \omega_{21}}{2}\sigma_3,
  \label{eq:h0}
\end{equation}
where $\hbar \omega_{21}$ is the energy difference between the two
states, $\sigma_i$ are the Pauli matrices ($i=1,2,3$) and $I$ is the
identity matrix. We will consider two cases for the external
potential: $V(\hat{z})=mg\hat{z}-m\gamma \hat{z}^2/2$
i.e. $V(\hat{z})$ at most quadratic in $\hat{z}$, and
$V(\hat{z})=mg\hat{z}+{\cal V}(\hat{z})$ where ${\cal V}(\hat{z})$ is
sufficiently weak to be treated as a small perturbation.  We assume
that the laser fields are classical, so the non-commuting
operators are only $\hat{z}$ and $\hat{p}$. To alleviate the notation,
henceforth we drop the hat from $\hat{z}, \hat{p}$ 
and their functions.

We 
consider a Kasevich-Chu type interferometer
\cite{Kasevich_1991}, where a sequence of three pulses
$\pi/2-\pi-\pi/2$ of temporal length $\tau,2\tau,\tau$ respectively
are separated by two free evolution intervals of length $T-2\tau$ so
that the total duration of the interferometric sequence is $2T$. We
remark that different sequences of pulses can also be considered \cite{Dubetsky_2006,Cadoret_2016}.
In present day interferometers the orders of magnitude of $\tau$ and
$T$ are $10^{-5} \mathrm{~s}$ and 1 s, respectively. We will also
assume $\tau_s \sim 10^{-4}$~s. 

In order to keep the optical field in
resonance with the atoms during their free fall, a phase-continuous,
linear frequency chirp on the laser fields partially compensates the
Doppler effect. Thus, the phase $\phi_L$ can be written as a function
of position and time as

\begin{equation}
  \phi_L(z,t) = \omega_0 t + \frac{\alpha t^2}{2}-kz.
    \label{eq:philas}
\end{equation}

Here $\omega_0$ is the frequency difference between the two Raman
beams, $k$ is the sum of the Raman beams wavenumbers and $\alpha$ is
the chirp rate. We make the simplifying assumption that $k$ is
constant in time (which is the case if both Raman beams are frequency
chirped in opposite directions), and we neglect any effect due to the
finite speed of light. A recent discussion on this last point is found in \cite{Tan_2016}.

Since in the following we will frequently use unitary transformations,
we recall that, under a generic unitary transformation
$\ket{\psi '}= U(t) \ket{\psi}$, the Hamiltonian transforms as
\begin{equation}
  H'(t) = U(t) H U^\dagger(t) + i \hbar (\partial_t U )  U^\dagger(t).
  \label{eq:uhu}
\end{equation}

The time-evolution operator over the generic time interval $[t_1, t_2]$ obeys the differential equation 
\begin{equation}
  i\hbar \partial_{t_2} {\cal U}(t_2,t_1)=H(t_2) {\cal U}(t_2,t_1),
\end{equation}
with the boundary condition ${\cal U}(t_1,t_1)=I$, whose general solution is the well-known time-ordered exponential 
\begin{equation}
  {\cal U}(t_2,t_1)={\cal T} \exp\left(-\frac{i}{\hbar}\int_{t_1}^{t_2} H(t')\,dt' \right), 
\end{equation}
usually calculated through the Dyson series. 
We will use instead an alternative expansion of ${\cal U}(t_2,t_1)$
called the Magnus expansion \cite{Blanes_2009,Blanes_2010}, for which
${\cal U}(t_2,t_1)$ is written as the exponential of a series

\begin{equation}
  {\cal U}(t_2,t_1)=\exp\left(\sum_{n=1}^{+\infty}M_n(t_2,t_1)\right).
  \label{eq:magser} 
\end{equation}

Differently from the Dyson series, the Magnus expansion preserves the
unitarity of ${\cal U}(t_2,t_1)$ at any order but, as a drawback, it
requires an operator exponentiation. A summary on the Magnus expansion
is found in App.~\ref{app:formulas}.

Under the generic time-dependent unitary transformation described above, the evolution operator is also transformed:
\begin{equation}
    {\cal U}'(t_2,t_1) = U(t_2) {\cal U}(t_2, t_1) U^\dagger(t_1) . 
\end{equation}

Following \cite{Marzlin_1996}, the time-dependent phase $\phi_L(z,t)$
is eliminated by means of the unitary transformation generated by
$U_3(t)=\exp[i \sigma_3 \phi_L(z,t)/2]$ (the index 3 indicates that
the exponent is proportional to $\sigma_3$).  After adopting the
Rotating Wave Approximation (RWA) \cite{Walls} to cancel the terms
oscillating as $\exp[i2 \phi_L(z,t)]$, the Hamiltonian transformed
under $U_3(t)$ reads

\begin{equation}
  \begin{split}
    H^{\mathrm{I}}(t)=& U_3(t)HU_3^\dagger(t)-\frac{\hbar \partial_t\phi_L(z,t)}{2}\sigma_3\\ 
    =&-\frac{\hbar\Omega(t)}{2}\sigma_1-\frac{\hbar}{2}\delta(t) \sigma_3\\
    &+\left(\frac{p^2}{2m}+\frac{\hbar^2 k^2}{8 m}+V(z)\right )I,    
  \end{split}
  \label{eq:h1}
\end{equation} 
where $\delta(t)$ is defined as the Doppler-shifted detuning

\begin{equation}
  \delta(t)=\Delta(t)+\frac{pk}{m},
  \label{eq:deltadef}
\end{equation}
with $\Delta(t) \equiv \omega(t)-\omega_{21}$ and
$\omega(t)\equiv \omega_0+\alpha t$.  Note that the transformed
momentum is $U_3(t) pI U_3^\dagger(t)=pI-\hbar k\sigma_3/2$ so the
transformation adds $\hbar k/2$ to, and subtracts $\hbar k/2$ from, the
momentum of states $\ket{1}$ and $\ket{2}$, respectively: this is equivalent
to a translation of the classical upper and lower trajectories on the mean
path, i.e. the trajectory with average momentum after the first
beamsplitter pulse, as shown in Fig.~\ref{fig:paths}.

An additional unitary transformation will eliminate the term
proportional to $I$ in $H^{\mathrm{I}}(t)$, which is equivalent
to moving to a reference frame in free fall. This operation is straightforward
if $V(z)$ is at most quadratic in $z$, otherwise we must apply
perturbation theory and assume that the potential is the sum of a
large linear part and a small term. We will consider the two cases
separately.

\section{\label{sec:quadPot} Quadratic potential}
We discuss the well studied case of a quadratic potential to
illustrate our method and derive with it well-known results.

For the Earth's gravitational field we use the second order potential
$V(z)=mgz-m\gamma z^2/2$, define
$H_\gamma= (p^2/2m+mgz -m\gamma z^2/2)I$ and apply the unitary
transformation $U_0(t)=\exp(i H_\gamma t/\hbar)$ to $H^{\mathrm{I}}$. Such
transformation changes the reference system to the freely falling one,
which is commonly adopted to describe the experiments in
weightlessness \cite{Becker_2018,Barrett_2016}. The result is
\begin{equation}
  H^{\mathrm{II}}(t)=-\frac{\hbar}{2}\left[\Omega(t)\sigma_1+\delta(t)\sigma_3\right],
  \label{eq:h2}
\end{equation}
where the momentum $p$ in $\delta(t)$ is now replaced by $p(t)$,
i.e. the momentum time-evolved according to the Heisenberg
representation with Hamiltonian $H_\gamma$
\begin{equation}
    p(t)=\displaystyle{p \cosh \sqrt \gamma t+mz\sqrt \gamma \sinh
      \sqrt \gamma t-\frac{m g \sinh \sqrt \gamma t}{\sqrt \gamma}}.
  \label{eq:pt}
\end{equation}
As expected, this expression coincides at $t=0$ with the
time-independent $p$ operator. Similarly, for the following we define
the time-evolved operator $z(t)$:
\begin{equation}
    z(t)=\displaystyle{z \cosh \sqrt \gamma t+p \frac{\sinh \sqrt \gamma t}{m\sqrt \gamma}+\frac{g(1-\cosh \sqrt{\gamma}t)}{\gamma}}.\\
  \label{eq:zt}
\end{equation}

In the case of the Earth's gradient
($\gamma \simeq 3 \times 10^{-6}\mathrm{s}^{-2}$) and present day
interferometers ($T \simeq 1 \mathrm{~s}$), we have
$2T \sqrt \gamma \ll 1$; Eqs.~(\ref{eq:pt}, \ref{eq:zt}) can then be expanded in
series up to the second order in $\sqrt \gamma t$ and, keeping only terms at
most linear in $\gamma$, one obtains a simpler approximate expression
for $\delta(t)$

\begin{equation}
\begin{split}
  \delta(t)\simeq &\Delta(0) + \frac{k p}{m} \left(1+\frac{\gamma t^2}{2}\right) 
  -(kg-\alpha)t\\
  &+k \gamma t \left(z - \frac{gt^2}{6} \right).
  \end{split}
\label{eq:deltaapprox}
\end{equation}
The expression above shows that, when $\alpha=kg$, the dominant time
dependent term in $\delta(t)$ is canceled and
$\delta(t)\simeq \delta(0)$, which is equivalent to the atoms seeing a
constant laser phase in their free fall. We remark that now
$[\delta(t),\delta(t')]\equiv ic_{\delta,\delta}(t,t')$ is a c--number
with $c_{\delta,\delta}(t,t')$ given by
\begin{equation}
  c_{\delta,\delta}(t,t')= k v_r\sqrt\gamma\sinh \sqrt \gamma (t-t') \simeq k v_r\gamma (t-t'),
  \label{eq:cdd}
\end{equation}
where we have defined the recoil velocity $v_r\equiv\hbar k/m$.

We seek to separate the effect of the free evolution from that of the
interferometer pulses. In this respect, the Hamiltonian of
Eq.~(\ref{eq:h2}) is still unsatisfactory: while the term proportional
to $\Omega(t)$ vanishes during the free evolution, its temporal
integral, i.e. the corresponding accumulated phase, cannot be
neglected since the pulses have an area $\sim\pi$. Therefore, we
define a third unitary transformation
\begin{align}
    U_1(t) & \equiv \exp[-i\phi_1(t)\sigma_1/2] ,\\
       \phi_1(t) & \equiv \int_0^t \Omega(t')\,dt' , 
\end{align}
which leads to the Hamiltonian
\begin{equation}
  H^{\mathrm{III}}(t)=\frac{\hbar}{2}\delta(t)\left[\sin \phi_1(t)\sigma_2-\cos \phi_1(t)\sigma_3\right].
  \label{eq:h3}
\end{equation}

We will see in the following that the Hamiltonian in Eq.~(\ref{eq:h3}) has the required form, i.e. the sum of a dominant term,
$ H_L^{\mathrm{III}}(t)$ proportional to $\sin\phi_1(t)\sigma_2$, plus a
small term, $H_S^{\mathrm{III}}(t)$ proportional to
$\cos\phi_1(t) \sigma_3$, which vanishes during the free evolution for
pulses with ideal area. Later we will refer to them as `ideal pulses'.

\subsection{\label{ss:happrox}Approximate solution}

\begin{figure}[t!]
  \includegraphics[scale=1]{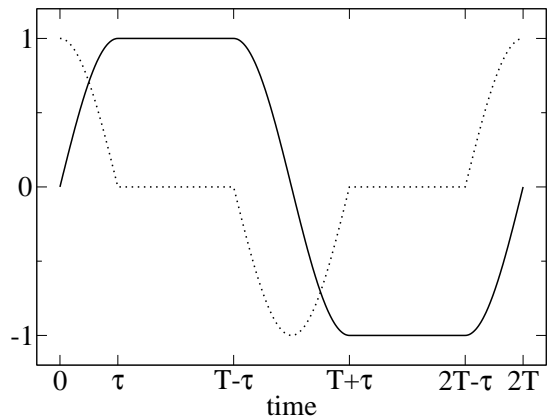}    
  \caption{\label{fig:sfun}Plot of the functions $\sin \phi_1(t)$
    (continuous line) and $\cos \phi_1(t)$ (dotted line) for ideal
    rectangular pulses in a Mach-Zehnder interferometer. The two functions
    are formed either by sinusoidal functions or horizontal lines. In
    this figure $\eta=\tau/T$ is 0.25 for clarity; typical
    experimental values for $\eta$ are in the $10^{-4} \sim 10^{-5}$
    range.}
\end{figure}

We aim to evaluate the transition probability for an atom in the
initial internal state $\ket{1}$ to exit the interferometer in
$\ket{2}$.

As a preliminary step, we neglect $H_S^{\mathrm{III}}(t)$: as shown in
Fig.~\ref{fig:sfun}, for ideal pulses $\cos \phi_1(t)=0$ during the
free evolution, thus $H_S^{\mathrm{III}}(t)$ is a perturbation acting
only during the interferometric pulses. We will evaluate the
corrections to this approximation in Sec. \ref{ss:hoc}.

To evaluate the probability amplitude we need the off-diagonal matrix
element of the evolution operator from $t=0$ to $t=2T$, for which we
revert to the Magnus expansion earlier introduced. As shown in
App.~\ref{app:formulas}, since in the present approximation
$[H(t),H(t')]$ is a c--number, the Magnus series terminates at
$M_2$. Defining
\begin{equation}
  \begin{cases}
    \phi_2(t) \equiv \displaystyle{\int_0^t \delta(t') \sin \phi_1(t')\,dt',}\\
    \\
    \psi_2(t) \equiv \displaystyle{\frac{1}{8} \int_0^t dt' \int_0^{t'}
      c_{\delta,\delta}(t',t'') \sin \phi_1(t') \sin \phi_1(t'')
      \,dt''},
  \end{cases}
  \label{eq:phipsi}
\end{equation}
we have $M_1=-i\phi_2\sigma_2/2$ and $M_2=-i\psi_2I$ (later, to simplify the notation, we will omit the temporal arguments when $t=2T$) and
\begin{equation}
   {\cal U}_L^{\mathrm{III}}(2T,0)=
   \displaystyle{\exp(-i\psi_2)\exp\left(-i\frac{\phi_2}{2}\sigma_2\right)},  
   \label{eq:U3}
 \end{equation}
where the subscript $L$ means that we consider only $H_L^{\mathrm{III}}(t)$.

The evolution operator in the mean path frame then reads
 ${\cal U}_L^{\mathrm{II}}(2T,0)=U_1^\dagger(2T){\cal U}_L^{\mathrm{III}}(2T,0)$.
 Since
 $U_1^\dagger(2T)=\exp[i\phi_1\sigma_1/2]$, the transition probability $P_{21}$
 from $\ket{1}$ to $\ket{2}$ at the output of the interferometric sequence can
 be evaluated directly:
\begin{align}
\label{eq:prob21}
  P_{21} & = |\bra{2}  \exp (i\phi_1 \sigma_1/2) \exp(-i\psi_2)\exp(-i \phi_2 \sigma_2 /2) \ket{1}|^2 \nonumber \\
   & = \frac{1}{2}\left( 1-\cos\phi_1 \cos\phi_2 \right),
\end{align}
where the 
internal states $|n\rangle$, with $n=1,2$, are evaluated in
the reference frame $\mathrm{II}$, i.e.
$|n(t)\rangle=U_0(t) U_3(t)|n\rangle$.

Ideally, the total pulse area $\phi_1$ in Eq.~(\ref{eq:prob21}) is
equal to $2\pi$ and the contrast $\cos\phi_1$ equal to 1. In case of
slightly imperfect pulses $\phi_1=2\pi+\delta \phi_1$, the effect of
$\delta \phi_1$ is just a contrast reduction of the interference
fringes.

Assuming ideal, rectangular pulses, it is simple to obtain a closed
form expression for $\phi_2$ from Eqs.~(\ref{eq:deltadef}, \ref{eq:pt}, \ref{eq:zt}).
Here we report only an approximate expression using Eq.~(\ref{eq:deltaapprox}),
keeping only terms up to the first order
in the small parameter $\eta=\tau/T$. This expression depends only on
the area, not on the actual shape, of the pulses:
\begin{widetext}
  \begin{equation}
    \phi_2=T^2(kg-\alpha-k\gamma z_0)\left(1-\frac{2\pi-4}{\pi}\eta
    \right)-k\gamma T^3\left[v_m\left(1-\frac{2\pi-4}{\pi}\eta\right)-gT\left(\frac{7}{12}-\frac{4\pi-8}{3\pi}\eta \right)\right],   
    \label{eq:phi2expanded}
  \end{equation}  
\end{widetext}
where we have used 
$p_0/m=v_0+v_r/2=v_m$ for the motion on the mean path.  We notice that some numerical coefficients in this formula do not agree with those
in Eq.~(40) of Ref.~\cite{Li_2015}.

The explicit inclusion of $z$ shows that for a gradiometer where two clouds
with the same initial velocity are separated by a distance
$d$ the differential phase shift is simply

\begin{equation}
  \Delta \phi_2=-k\gamma dT^2\left(1-\frac{2\pi-4}{\pi}\eta \right).
\label{eq:phi2grad}
\end{equation}  

The formula for $P_{21}$ can be easily understood by
noting that $H_L^{\mathrm{III}}(t)$ is diagonalized by the time
independent eigenvectors
  \begin{equation}
    \ket{\pm}=\frac{\ket{1} \pm i\ket{2}}{\sqrt{2}} \Rightarrow \ket{1}=
    \frac{\ket{+}+\ket{-}}{\sqrt{2}}, \ket{2}=
    \frac{\ket{+}-\ket{-}}{i\sqrt{2}}
    \label{eq:eigen3}
\end{equation}
with the time dependent eigenvalues
\begin{equation}
  E^{\pm}(t)=\pm \frac{\hbar \delta(t) \sin\phi_1(t)}{2}.
  \label{eq:pmeigenvals}
\end{equation}
One must have, due to the interference between $\ket{+}$ and $\ket{-}$,
\begin{equation}
  P_{21}= \frac{1}{4}\left|1-\exp\left(\frac{i}{\hbar}\int_0^{2T}E^+(t)-E^-(t)\,dt\right)\right|^2,
  \label{eq:prob21bis}
\end{equation}
which is equivalent to Eq.~(\ref{eq:prob21}) for ideal pulses. This is
analogous to observing the Rabi oscillations in the dressed
atom picture \cite{Cohen_2008}.

\subsection{\label{ss:hoc}Effect of the full Hamiltonian}
To take into account $H_S^{\mathrm{III}}(t)$ we apply another
unitary transformation:
\begin{equation}
U_2=\exp(i\psi_2(t))\exp(i\phi_2(t)\sigma_2/2).
\end{equation}
Since $U_2^\dagger={\cal U}_L^{\mathrm{III}}(t,0)$, this unitary transformation is
just the interaction representation with respect to
$H_L^{\mathrm{III}}(t)$.

The new Hamiltonian $H^{\mathrm{IV}}(t)$ in the interaction representation is the
transform of $H_S^{\mathrm{III}}(t)$,
\begin{equation*}
    H^{\mathrm{IV}}(t)=-\frac{\hbar}{2}U_2 \delta(t)[\cos \phi_1(t)\sigma_3] U_2^\dagger,
\end{equation*}  
which can be evaluated using Eq.~(\ref{eq:paulievolop}), by letting
$B=\delta$ and $A=\phi_2/2$, as

\begin{equation}
  \begin{split}
    H^{\mathrm{IV}}(t)&= \frac{\hbar}{4} \left[ \{ \delta(t), \sin \phi_2(t)\}\sigma_1- \{\delta(t), \cos \phi_2(t)\}\sigma_3\right]\\
    &\times \cos \phi_1(t), \\
  \end{split}
    \label{eq:h4}
\end{equation}  
where $\{\,,\}$ denotes the anticommutator.

Using the general identity to transform the evolution operators, we obtain
\begin{equation}
  \begin{split}
    {\cal U}^{\mathrm{IV}}(2T, 0)&= U_2(2T) {\cal U}^{\mathrm{III}}(2T,0) \\
    &= [{\cal U}_L^{\mathrm{III}}(2T,0)]^\dagger {\cal U}^{\mathrm{III}}(2T,0) 
  \end{split}
\end{equation}
or, equivalently:
\begin{equation}
  {\cal U}^{\mathrm{III}}(2T,0)= {\cal U}_L^{\mathrm{III}}(2T,0) {\cal U}^{\mathrm{IV}}(2T, 0).
\end{equation}
Therefore, ${\cal U}^{\mathrm{IV}}(2T,0)$ is the multiplicative correction
sought to take into account $H_S^{\mathrm{III}}(t)$.

Under certain conditions, ${\cal U}^{\mathrm{IV}}(2T,0)$ is
easily evaluated: if the pulses are ideal, during the
free evolution we have $H^{\mathrm{IV}}(t)=0$ and ${\cal U}^{\mathrm{IV}}=I$ and thus,
\begin{equation}
  \begin{split}
    {\cal U}^{\mathrm{IV}}(2T, 0) &= {\cal U}^{\mathrm{IV}}(2T, 2T-\tau)
    {\cal U}^{\mathrm{IV}}(T+\tau, T-\tau)\\
    &\times {\cal U}^{\mathrm{IV}}(\tau, 0).
  \end{split}
  \label{eq:P1}
\end{equation}
If the pulses are short, $\delta(t)$ can be
considered constant during the pulses and we find
\begin{equation}
  \begin{split}
    {\cal U}^{\mathrm{IV}}(2T, 0)=&\exp[-i\theta(2T)\hat n(\phi_2(2T))\cdot\vec{\sigma}]\\
    &\times \exp[2i\theta(T)\hat n(\phi_2(T))\cdot\vec{\sigma}]\\
    &\times \exp[-i\theta(0)\hat n(\phi_2(0))\cdot\vec{\sigma}],
  \end{split}
  \label{eq:P2}
\end{equation}
where we have defined $\theta(t) \equiv \tau \delta(t)/2$ and
$\hat n(\phi_2(t))\equiv(\sin\phi_2(t),0,-\cos\phi_2(t))$. To alleviate
the notation, we have written the half-anticommutators as products,
e.g. $\theta(t) \hat n_i(\phi_2(t))$ for
$\{\theta(t), \hat n_i(\phi_2(t))\}/2$.

Clearly if $|\theta(t)|\ll 1$ then ${\cal U}^{\mathrm{IV}}(2T,0) \simeq I$.
The effect of the correction ${\cal U}^{\mathrm{IV}}(2T,0)$ is to reduce the
contrast in the interference fringes and to introduce an additional
phase shift $\delta\phi_2$ with respect to Eq.~(\ref{eq:prob21}).
Such a phase shift can be evaluated explicitly by applying repeatedly
the product rule for exponential of Pauli vectors (See
App.~\ref{app:formulas}) only if we assume $\gamma=0$ so all the
commutators involving $\theta(t)$ and $\phi_2(t)$ are zero. Here we
report only the approximate result when $|\theta(t)| \ll 1$ by
expanding $P_{21}$ to leading-order terms in $\theta(t)$ and
$\phi_2(\tau)$, which is of the same order as $\theta(\tau)$. After
some algebra we obtain
\begin{equation}
  \delta \phi_2=-4\theta^2(T)\sin 2\phi_2(T)+O(\theta^3).
  \label{eq:pt4}
\end{equation}

This is one of the main results of our analysis, showing that the
interferometric phase shift carries an additional contribution due to
the evolution during the laser pulses, actually dominated by the
central $\pi$ pulse at time $t=T$. However, this contribution is
easily washed out by averaging over the velocity distribution of the sample: in
typical experimental conditions the width of the velocity
distribution is inversely proportional to the duration $\tau_s$ of the
selection pulse, and $\tau,\tau_s, T$ obey to $\tau < \tau_s \ll T$;
thus, we have simultaneously
$|\theta(t)|^2 \sim \tau^2/\tau^2_s \ll 1$ and
$\phi_2(T) \sim T/\tau_s \gg 1$ with $\phi_2(T)$ varying rapidly with
the initial detuning $\delta(0)$. As a consequence, $\delta \phi_2$
averages to zero over the atomic sample and the phase shift evaluated
in Eq.~(\ref{eq:phi2expanded}) still holds.

The effect of non ideal pulses has been considered in
\cite{Bonnin_2015}, for rectangular pulses, using the SF formalism,
equivalent to our treatment in Sec. \ref{ss:happrox}. There the terms
proportional to $\delta(0)$ are retained and not assumed to cancel
after the average over the initial velocity distribution.

\subsection{\label{ss:loc}Loss of contrast}
In general, in a nonlinear potential, the end points of the upper and
lower paths do not coincide. The loss of contrast induced by this
effect and the strategies to mitigate it are discussed in
\cite{Roura_2014,Roura_2017} and experimentally implemented in
\cite{Damico_2017,Overstreet_2018}. Here we derive in our formalism
the conditions to achieve high contrast in the case of a constant
gradient, in order to extend them later to an arbitrary weak
perturbing potential.

We start by evaluating the operators $z(t), p(t)$ after the unitary
transformation generated by $H_L^{\mathrm{III}}$, using
Eq.~(\ref{eq:hadamard}) in App.~\ref{app:formulas}, at time $t=2T$
obtaining
\begin{equation}
  \begin{split}
    z(2T)&=z_m(2T)I+\frac{i\sigma_2}{2}[\phi_2(2T),z_m(2T)]\\
    &= z_m(2T)I+\frac{\sigma_2v_r}{2}\int_0^{2T}\!\!\sin\phi_1(t)\,\cosh \sqrt{\gamma}(2T-t) \,dt\\
    &\simeq z_m(2T)I+\frac{\sigma_2v_r\gamma T^3}{2},
  \end{split}
\end{equation}
and, similarly, 
\begin{equation}
  p(2T)\simeq p_m(2T)I+\frac{\sigma_2mv_r\gamma T^2}{2}.
\end{equation}
The eigenvectors of both operators are again $\ket{\pm}$.  The
separation in position and momentum is given by the difference between
the eigenvalues, i.e. $\Delta z(2T)=v_r\gamma T^3$ and
$\Delta p(2T)=m v_r\gamma T^2$.

In \cite{Roura_2014} it is shown that the condition
$\Delta z(2T)=0,\Delta p(2T)=0$ at the end of an interferometric
sequence ensure high contrast independently from the detection
time. More generally, high contrast is obtained when
$\Delta z(2T)-\Delta t_d\Delta p(2T)/m=0$, where $\Delta t_d$ is the
time interval between the last pulse and detection. By slightly
changing the duration of the second free evolution period it is
possible to fulfill only the latter condition.

A better strategy, suggested in \cite{Roura_2017} and demonstrated in
\cite{Damico_2017,Overstreet_2018}, is to change the momentum of the
Raman beams by an amount $\delta k$ at the $\pi$ pulse. In this way
$v_r$ is changed by an amount $\delta v_r=2\hbar \delta k/m$ during
the second free evolution: by choosing $\delta v_r/v_r=-\gamma T^2$,
$\Delta z(2T)$ vanishes while the effect of $\Delta p(2T)$ is
negligible. Now however in Eq.~(\ref{eq:philas}) we have $k=k(t)$ and,
due to the time derivative in Eq.~(\ref{eq:uhu}), an extra term
appears in the Hamiltonian, providing a momentum kick at the $\pi$
pulse that exactly compensates $\Delta p(2T)$.  The key to the
possibility of compensating simultaneously $\Delta z(2T)$ and
$\Delta p(2T)$ lies in the relation $m \Delta z(2T)/ \Delta p(2T)=T$.

We will show in Sec. \ref{ss:locpert} that this condition does not hold in
general if $V(z)$ is more than quadratic.

\subsection{\label{ss:cfr}Comparison with previous results}

Here we show that Eq.~(\ref{eq:phi2expanded}) is consistent with
previous literature.

Except for a sign, $\sin \phi_1(t)$ coincides with the SF
introduced in \cite{Cheinet_2008} for rectangular pulses and it is
immediately applicable to more general cases i.e. Gaussian or
imperfect pulses. Note that even if the SF neglects
$H^{\mathrm{III}}_S(t)$ in Eq.~(\ref{eq:U3}), the phase
shift averaged over the initial atomic velocity distribution is
correct as shown in Eq.~(\ref{eq:pt4}).

If we use the expression for $\delta(t)$ given in
Eq.~(\ref{eq:deltadef}) and, moving to the expectation values, apply
the Ehrenfest's theorem replacing $p/m$ with $\dot z$, we can
integrate by parts the first expression in Eq.~(\ref{eq:phipsi}) in
the case of ideal rectangular pulses of negligible duration

\begin{equation}
  \begin{split}
    \phi_2&=-\int_0^{2T} [\phi_L(t)+kz_m(t)]\Omega(t) \cos \phi_1(t)\,dt\\
    &\simeq -D_2[\phi_L]-kD_2[z_m],
    \label{eq:5pts}
  \end{split}
\end{equation}
where $\phi_L(t)$ is the primitive of $\Delta(t)$ and, to simplify the
notation, we have defined $D_2[f]\equiv f(2T)-2f(T)+f(0)$.
The boundary term of the integration by parts vanishes, for ideal pulses, as
$\sin\phi_1(2T)=\sin\phi_1(0)=0$.  Note that in $D_2[\phi_L]$
the terms constant and linear in $t$ disappear so
$-D_2[\phi_L]=\alpha T^2$ while, since $z(t)$ and $p(t)$ are
linear in $z$ and $p$ in Eqs.~(\ref{eq:pt}, \ref{eq:zt}), then
$2z_m(t)=z_u(t)+z_l(t)$ so
\begin{equation}
  D_2[z_m]=\frac{z_u(2T)+z_l(2T)}{2}-(z_u(T)+z_l(T))+z_m(0),
\end{equation}
which is the result given in \cite{Antoine_2003}.

Next, we compare Eq.~(\ref{eq:5pts}) with the path integral
prescription, as described, for example, in \cite{Bongs_2006}, where
the phase shift is evaluated as the sum of three terms,
$\delta \phi_L+\delta \phi_p+\delta \phi_s$.  The `laser' term
$\delta \phi_L$ is given by
\begin{equation}
  \begin{split}
    \delta \phi_L&=\phi_L(0)+kz_m(0)-2\phi_L(T)-k[z_u(T)+z_l(T)]\\
    &+\phi_L(2T)+kz_l(2T),    
  \end{split}
\end{equation}
where $\phi_L(t)\equiv \phi_L(0,t)$ and $\phi_L(z,t)$ is given by
Eq.~(\ref{eq:philas}).

The `propagation' term $\delta \phi_p$ is given by

\begin{equation}
  \delta \phi_p=\frac{1}{\hbar}\left(\int_u {\cal L} dt-\int_d {\cal L} dt\right)
  \equiv \frac{1}{\hbar}\oint_{cp} {\cal L} dt,
\end{equation}
where the two integrals are along the upper and lower classical paths
and ${\cal L}$ is the Lagrangian. To simplify the notation the
difference of the two integrals is denoted as a circulation integral
along the classical path $cp$ even if $cp$ is not closed.

In case of a quadratic potential it is easy to see that the kinetic
and the potential energies give equal contributions to the integral
so $\delta \phi_p=0$.

Finally the `separation' term is defined as

\begin{equation}
  \delta \phi_s=\frac{k(z_u(2T)-z_l(2T))}{2},
\end{equation}
where we have taken into account that the average momentum of the two
states in an output channel must be measured on the mean path. Clearly
the path integral prescription gives the same result as
Eq.~(\ref{eq:5pts}).
  
\begin{figure}[t!]
  \includegraphics[scale=1]{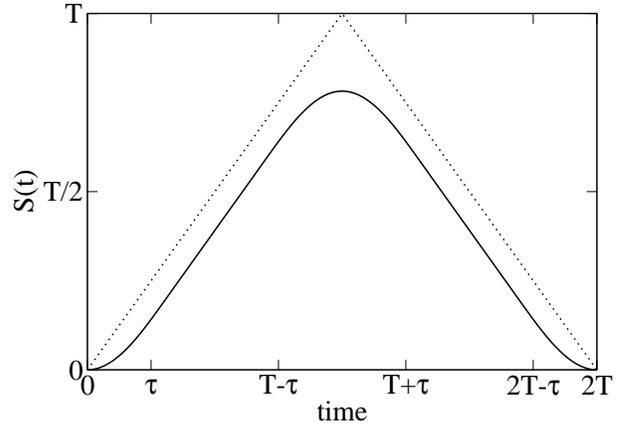}    
  \caption{\label{fig:S} Plot of the function $S(t)=\int_0^t \sin \phi_1(t')\,dt'$
    for $\eta=0.25$ (continuous line) and $\eta=0$ (dotted line) for square pulses.}
\end{figure}

Another possibility to evaluate $\phi_2$ involves integrating by
parts the term $kp(t)/m$ in $\delta(t)$ in the other order,
replacing $\dot p$ with $-\partial_z V$ and obtaining, in the same
hypothesis as above, the contribution to $\phi_2$ due to $V$,
$\phi_2^V$ as

\begin{equation}
    \phi_2^V \simeq \frac{v_r}{\hbar}\int_0^{2T}S(t) \partial_z V(t)\,dt ,
  \label{eq:accmp}
\end{equation}
where $S(t)$ is the primitive of $\sin \phi_1(t)$, see
Fig.~(\ref{fig:S}).  Noting that for a quadratic potential
$V(z+\Delta z)-V(z-\Delta z)= 2\Delta z \partial_z V(z)$ we can write

\begin{equation}
  \phi_2^V=\frac{1}{\hbar}\oint_c V(t)\,dt,
  \label{eq:phivar}
\end{equation}
where the closed path $c$ is delimited by $z_m(t) \pm v_r S(t)/2$. We
can also express $\phi_2^V$ as the difference of two integrals on the
upper and lower classical paths by taking
$z_u(t)-z_l(t) \equiv S(t)v_r+\delta z(t)$ as a definition of
$\delta z(t)$ to obtain

\begin{equation}
  \begin{split}
    \phi_2^V=&\frac{1}{\hbar}\oint_{cp} V(t)\,dt - \frac{1}{\hbar}\int_0^{2T}\delta z(t) \partial_z V\,dt\\
    =&\frac{1}{\hbar}\oint_{cp} V(t)\,dt +\frac{p(2T)[z_u(2T)-z_l(2T)]}{\hbar}\\
    &-\frac{1}{\hbar}\int_0^{2T} p(t) \delta v(t) \,dt,\\
    \label{eq:propsep}
  \end{split}
\end{equation}
where $\delta v=\delta \dot z$. Note that $\delta v=0$ during the free
evolution. Here the phase shift can be interpreted as propagation
term depending only on the potential, a separation term and finally
a term that contains the correction for the finite duration of the
pulses.

\section{\label{s:ppot}Perturbative potential}
If the potential $V(z)=mgz+{\cal V}(z)$ is more than quadratic, a
solution for the Heisenberg equations for $z$ and $p$ is in general
not known, so it is not possible to transform to the free fall
reference frame. Except for some special
choice of $V(z)$, in general $\delta(t)$ will not be linear in both
$z$ and $p$ so $[H^{}(t),H^{}(t')]$ will not be a c--number,
preventing an exact calculation of ${\cal U}^{\mathrm{III}}$ as in
Eq.~(\ref{eq:U3}).

Here we adopt a perturbative approach that works when ${\cal V}(z)$ is
small, in a sense that will be defined precisely later. In this way, we
get an approximate result even for a purely quadratic potential, with 
the advantage of a much simpler algebra. 

We use $U_0(t)$ generated by 
\begin{equation}
  H_m=\left(\frac{p^2}{2m}+mgz +\frac{\hbar^2k^2}{8m}\right)I
\end{equation}
obtaining 
\begin{equation}
  H^{\mathrm{II}}(t)=-\frac{\hbar}{2}\left[\Omega(t)\sigma_1+\delta(t)\sigma_3\right]+{\cal V}(z)I,
  \label{eq:h2p}  
\end{equation}
where in $H^{\mathrm{II}}(t)$ now $p(t)$ and $z(t)$ are given by 
Eqs.~(\ref{eq:pt}, \ref{eq:zt}) when letting $\gamma \rightarrow 0$. This
corresponds to a reference frame falling with constant acceleration
$g$. 

In the same way $\delta(t)$ is given by Eq.~(\ref{eq:deltaapprox})
with $\gamma=0$ so $c_{\delta,\delta} (t,t')=0$.

The evaluation of $H^{\mathrm{III}}(t)$ is straightforward since
$[U_1,{\cal V}(z)I]=0$:

\begin{equation}
  H^{\mathrm{III}}(t)=\frac{\hbar}{2}\delta(t)\left[\sin \phi_1(t)\sigma_2-\cos \phi_1(t)\sigma_3\right]+{\cal V}(z)I.
  \label{eq:h3p}
\end{equation}

Finally we apply the last unitary transformation $U_2$ as outlined in
Sec.~\ref{ss:hoc}. Note that, since now $\gamma=0$ all the commutators
not involving $z(t)$ are zero and we can write

\begin{equation}
  H^{\mathrm{IV}}(t)=H_\delta^{\mathrm{IV}}(t)+H_{\cal V}^{\mathrm{IV}}(t),
  \label{eq:h4p}
\end{equation}
where $H_\delta^{\mathrm{IV}}(t)$ is given, as in the harmonic potential case,
by Eq.~(\ref{eq:h4}) and
\begin{equation}
  H_{\cal V}^{\mathrm{IV}}(t)={\cal V}^+(z,v_rS/2)I+{\cal V}^-(z,v_rS/2)\sigma_2,
  \label{eq:h4pV}
\end{equation}
where we have defined
\begin{equation}
  {\cal V}^\pm(z,\Delta z) \equiv \frac{{\cal V}(z+\Delta z)\pm {\cal V}(z-\Delta z)}{2},
  \label{eq:vpm}
\end{equation}
and $H_{\cal V}^{\mathrm{IV}}(t)$ has been evaluating using
Eqs.~(\ref{eq:hadamard}, \ref{eq:funcom}).

\subsection{Approximate solution}
Here we evaluate again the time-evolution operator
${\cal U}^{\mathrm{IV}}(2T,0)$ neglecting the
$H^{\mathrm{IV}}_S(t)$ in $H_\delta^{\mathrm{IV}}(t)$. Later
we will take into account the full Hamiltonian.

We define ${\cal V}(z)$ as `small' when we can take
$[{\cal V}^\pm(t),{\cal V}^\pm(t')]= 0$ (see App.~\ref{app:approx})
and truncate the Magnus series for $H_{\cal V}^{\mathrm{IV}}(t)$ to the first
order: in this case the evolution operator is
${\cal U}^{\mathrm{IV}}(2T,0)=\exp[-i\epsilon_0I-i\epsilon_2\sigma_2 ]$ with
\begin{equation}
  \begin{cases}
    \displaystyle{\epsilon_0 =\frac{1}{\hbar}\int_0^{2T}{\cal V}^+(z,v_rS/2)\,dt'},\\
    \\
    \displaystyle{\epsilon_2 =\frac{1}{\hbar}\int_0^{2T}{\cal V}^-(z,v_rS/2)\,dt'}.\\
  \end{cases}
\label{eq:epsV}
\end{equation}

To evaluate the transition probability $P_{12}$, we need to transform back to
the previous reference frame, obtaining for the evolution operator
\begin{equation}
  {\cal U}^{\mathrm{III}}(2T,0)=\exp(-i\epsilon_0)\exp\left(-i\frac{\phi_2+2\epsilon_2}{2}\sigma_2 \right),
  \label{eq:U3V}
\end{equation}
where we have used the fact that, for ideal pulses, $\phi_2$
commutes with any analytic function of $z$ since
$[z(t),\phi_2(2T)]=iv_rS(2T)=0$,
and that, since $\cal V$ is small, we can let
$[\epsilon_0,\epsilon_2]=0$ and write
${\cal U}^{\mathrm{III}}(2T,0) = {\cal U}^{\mathrm{II}}(2T,0)$.

The term containing $\epsilon_0$ is an irrelevant phase factor, while
$2\epsilon_2$ is the additive phase shift to $\phi_2$ due to the
perturbing potential ${\cal V}$.  We can write $\epsilon_2$ using the
Taylor series for ${\cal V}(z_m)$ i.e. on the mean path
\begin{equation}
  2\epsilon_2=\frac{2}{\hbar}\sum_{n=0}^{\infty} \frac{v_r^{2n+1}}{(4n+2)!!}\int_0^{2T} S^{2n+1}(t)\partial_z^{2n+1}{\cal V}(z_m)\,dt.
  \label{eq:eps2ps}
\end{equation}
Note that the first term of the series above extends to $\cal V$ the
results obtained for a quadratic potential while higher order terms
are present only when $\partial_z^3 {\cal V} \ne 0$.

\subsection{Comparison with previous results}
The problem of non quadratic potentials has been discussed in
\cite{Roura_2014} locally solving for a quadratic potential but
assuming time varying values for $g$ and $\gamma$ along the atomic
trajectories.

The density matrix approach in the Wigner representation has been
adopted by Dubetsky in various papers, see i.e. \cite{Dubetsky_2006}
or, more recently, \cite{Dubetsky_2016}, on the mean path and also in
\cite{Roura_2014,Giese_2015} by considering the evolution along the
upper an lower paths and evaluating the relative interference
term. The equivalence with the path integral approach has
already been considered in \cite{Dubetsky_2017} so we
postpone a brief discussion on this subject to App.~\ref{app:wig}.

\subsection{\label{ss:locpert}Loss of contrast}
To evaluate $z(2T)$ after the unitary transformation
${\cal U}^{\rm{IV}}(2T,0)$ we note that $H_\delta^{\rm{IV}}$ does not
contribute at $t=2T$ for ideal pulses and the part proportional to $I$
in $H_{\cal V}^{\mathrm{IV}}$ has no effect on $\Delta z(2T)$; we need then
to evaluate only the commutator $[\epsilon_2,z(t)]$ with
$\epsilon_2$ from Eq.~(\ref{eq:epsV}).

From $[z(t),z(t')]= i\hbar (t'-t)/m$, Eq.~(\ref{eq:funcom}) leads to
\begin{equation}
  \Delta z(2T)=\frac{v_r}{m}\int_0^{2T}(2T-t)S(t) \partial_z^2{\cal V}\,dt,
  \label{eq:dzp}
\end{equation}
which generalizes the expression for $\Delta z(2T)$ obtained in the quadratic case.

For $p(t)$, with the help of $[p(t),p(t')]=0$ and $[z(t),p(t')]=i\hbar$, we obtain
\begin{equation}
  \Delta p(2T)=v_r\int_0^{2T}S(t) \partial_z^2{\cal V}\,dt.
  \label{eq:dzp}
\end{equation}

In general it is not possible to have
$m \Delta z(2T)/\Delta p(2T) \simeq T$ if
$\partial_z^3 {\cal V} \ne 0$ so the scheme suggested in
\cite{Roura_2017} is not extensible to arbitrary potentials but
compensates only the average gradient over the classical trajectory. A
straightforward modification, however, would be to use the change of
$k$ in the $\pi$ pulse to cancel $\Delta z(2T)$ and partially erase
$\Delta p(2T)$ and then change again $k$ at the last $\pi/2$ pulse to
complete the $\Delta p(2T)$ compensation.

\subsection{\label{ss:eofh}Effect of the full Hamiltonian}
We can evaluate the effect of $H_\delta^{\rm{IV}}(t)$ by applying a unitary
transformation that removes $H_{\cal V}^{\rm{IV}}(t)$ from
Eq.~(\ref{eq:h4p}) and, as in the quadratic case, obtain a resulting
Hamiltonian $H^V$ which is nonzero only during the pulses. The
evaluation of $H^V$ is straightforward if we make the approximation
$[{\cal V}^\pm(t),\delta(t')]=0$, justified in App.~\ref{app:approx}.
Analogously to the quadratic case, we obtain the new Hamiltonian
\begin{equation}
  H^V(t)=\frac{\hbar}{2}\delta(t)\cos \phi_1(t) \hat n(\phi_2(t)+2\epsilon_2(t))\cdot \vec \sigma ,
  \label{eq:h5p}
\end{equation}
which is the same as the one in Eq.~(\ref{eq:h4})
after the substitution $\phi_2(t) \rightarrow \phi_2(t)+2\epsilon_2(t)$.
Again we have assumed that all the operators are commuting so also
Eq.~(\ref{eq:P2}) and Eq.~(\ref{eq:pt4}) and the relative considerations
about sample averaging apply.

\section{\label{s:conclusions}Conclusions}
In a simple 1D model we have addressed the effects of the finite
duration of the interferometric pulses and of the presence of more than
quadratic perturbative potentials in the calculation of the phase
shift for atomic interferometers.

In the case of quadratic potentials, we have recovered the already known
interferometric phase shift, Eq.~(\ref{eq:phi2expanded}), for short
pulses, i.e. to first order in $\tau/T$.

We have also shown that the finite duration of the pulses is accurately
described by the SF method \cite{Cheinet_2008}: the
additional phase shift generated by the part of the Hamiltonian it neglects,
described by Eq.~(\ref{eq:pt4}), vanishes when averaged over a typical initial
velocity distribution of the atomic sample. Further, to take into account the
finite pulse duration in the path integral formalism, we have derived
Eq.~(\ref{eq:propsep}).

We have also shown how our formalism naturally describes the final
separation of the interferometer paths caused by the potential
curvature and causing a contrast reduction in the interferometric
fringes.

Finally, to lift the restriction of quadratic potentials we have
evaluated perturbatively the phase shift due to an arbitrary weak
potential, given by Eq.~(\ref{eq:eps2ps}) that generalizes a similar
result derived for quadratic potentials \cite{Storey_1994}.

\section{\label{s:acknowledgements}Acknowledgments}

We would like to thank B. Dubetsky for making available to us
preliminary versions of his manuscripts and for useful discussions,
B. Barrett, N. Gaaloul and G. Lamporesi for a careful reading of the
manuscript. A. B. acknowledges funding from Horizon 2020 QuantERA
ERA-NET -- Project TAIOL. M. P. acknowledges financial support from
LAPHIA--IdEx Bordeaux.

\appendix

\section{\label{app:formulas}Useful formulas}

Here we report for sake of completeness some useful formulas used in
the article.

We start with the first three terms of the Magnus expansion:

\begin{equation}
  \begin{cases}
    M_1(t,0)=\displaystyle{-\frac{i}{\hbar}\int_0^t\!\! H_1 \,dt_1},\\ 
    \\ 
    M_2(t,0)=\displaystyle{-\frac{1}{2\hbar^2}\int_0^t\!\! dt_1\!\! \int_0^{t_1}\!\! [H_1,H_2]\,dt_2}, \\              
    \\
    \begin{split}
      M_3(t,0)=\displaystyle{-\frac{i}{6\hbar^3}\int_0^t\!\! dt_1\!\! \int_0^{t_1}\!\! dt_2\!\! \int_0^{t_2}\!\!}&[H_1,[H_2,H_3]]\\
      +&[H_3,[H_2,H_1]]\,dt_3,
    \end{split}
  \end{cases}  
  \label{eq:magnus}
\end{equation}  
where $H_n$ is a shortened notation for $H(t_n)$. A recursion
formula for generating successive terms is known \cite{Blanes_2010}.

Another identity that we often used is
\begin{equation}
  \exp(\alpha A) B\exp(-\alpha A)=\sum_{n=0}^{+\infty} \frac{\alpha^n}{n!} \mathrm{ad}_A^n B,
  \label{eq:hadamard}
\end{equation}
where $\alpha$ is a complex number, $A,B$ are operators and
$\mathrm{ad}_A^n B$ is a nested commutator defined by recursion as
\begin{equation}
  \begin{cases}
    &\mathrm{ad}_A^n B = [A,\mathrm{ad}_A^{n-1}B] \qquad n>0,\\
    \\
    &\mathrm{ad}_A^0 B = B.\\
  \end{cases}
\end{equation}

When $A,B$ are Pauli matrices $\sigma_i,\sigma_j$, respectively,
with $i\ne j$ and $\alpha=i\theta$ it is easy to show that Eq. \ref{eq:hadamard}
becomes
\begin{equation}
  \exp(i\theta \sigma_i) \sigma_j \exp(-i\theta \sigma_i)=\sigma_j \cos 2\theta -\epsilon_{ijk}\sigma_k\sin 2 \theta.
  \label{eq:paulievol}
\end{equation}

If $A$ and $B$ are scalar operators for which $[A,[A,B]]=0$,
Eq.~(\ref{eq:paulievol}) can be generalized to
\begin{equation}
  \begin{split}
    \exp(iA \sigma_i) B \sigma_j \exp(-iA \sigma_i)=&\frac{\{B,\cos 2A\}}{2}\sigma_j\\
    &-\frac{\{B,\sin 2A\}}{2}\epsilon_{ijk}\sigma_k . \\
  \end{split}
  \label{eq:paulievolop}
\end{equation}

We also remind that

\begin{equation}
  \exp(i\vec n \cdot \vec \sigma) = I\cos n+i \vec \sigma \cdot \hat n \sin n,
  \label{eq:pauliexp}
\end{equation}
where $\vec n$ is a vector, $n$ its modulus and $\hat n=\vec n/n$ the
related unit vector. Note that, if $\vec n$ is a vector of operators,
Eq.~(\ref{eq:pauliexp}) holds only if $[n_i,n_j]=0$.

The product of two of these matrices is
\begin{equation}
  \begin{split}
    \exp(i\vec n \cdot \vec{\sigma}) \exp(i\vec m \cdot \vec{\sigma})=&I[\cos n \cos m -\hat n \cdot \hat m \sin n \sin m)]\\
    &+i[\hat n \sin n \cos m+\hat m \cos n \sin m)\\
    &-\hat n \times \hat m \sin n \sin m)]\cdot \vec \sigma ,      
  \end{split}
  \label{eq:pauliexpprod}
\end{equation}  
so it is of the same form as the two factors.

Another useful expression, if $[A,B]=c_{AB}$, where $c_{AB}$ is a
c--number, and $f$ and $g$ are analytic functions, is \cite{Transtrum_2005}

\begin{equation}
  [f(A),g(B)]=-\sum_{n=1}^{\infty}\frac{(-c_{AB})^n}{n!}\partial_A^n f(A) \partial_B^n g(B).
  \label{eq:funcom}
\end{equation}
\section{\label{app:approx}Perturbing potential approximations}
Here we discuss when the approximations involving the perturbing
potential, namely $[{\cal V}^\pm(t),{\cal V}^\pm(t')]\simeq 0$ and
$[{\cal V}^\pm(t),\delta(t')]\simeq 0$, assumed in Sec.~\ref{s:ppot},
are justified. We need to show that the commutators above are
negligible when compared with their anticommutators.

We start evaluating the following commutators when $\gamma=0$:
\begin{equation}
  \begin{cases}
    \displaystyle{[z(t),z(t')]=\frac{\hbar}{m}(t-t'),}\\
    \\
    [z(t),\delta(t')]=iv_r.\\
  \end{cases}
  \label{eq:zdeltacom}
\end{equation}

Note that the first commutator above defines a length scale
$z_0 \sim \sqrt{\hbar T/m}$ which, for heavy atoms, like Rb or Cs, and
$T\sim 1$~s is in the 25 $\mu$m range.

Both ${\cal V}^+$ and ${\cal V}^-$ can be approximated with
expressions evaluated on the mean path, ${\cal V}(z_m)$ and
$S v_r \partial_z{\cal V}(z_m)/2$ respectively. We can then apply
Eq.~(\ref{eq:funcom}) and Eq.~(\ref{eq:zdeltacom}). Since $z_0$ is
much smaller than the scale over which ${\cal V}$ is expected to vary
significantly, we can keep only the first nonzero term in the sum in
Eq.~(\ref{eq:funcom}) and note that, for example,
$[{\cal V}^+(t),{\cal V}^+(t')]$ is of the order of
$\delta{\cal V}^+(t)\delta{\cal V}^+(t')$ with $\delta{\cal V}^+$
being the increment of ${\cal V}$ over a distance of the order of
$z_0$.  Almost everywhere on the mean path then
${\cal V}^+(t){\cal V}^+(t') \gg \delta{\cal V}(t)\delta{\cal V}(t')$
holds. A similar argument can be applied to the other three
combinations of signes in $[{\cal V}^\pm(t),{\cal V}^\pm(t')]$.

For $[{\cal V}^\pm(t),\delta(t')]$, choosing $A=z(t)$ and
$B=\delta(t')$ in Eq.~(\ref{eq:funcom}), we need to show that
$|{\cal V}^\pm(t),\delta(t')| \gg |\partial_z {\cal V}^\pm(t)
v_r|$. Here we note that, in case of a sample of atoms that have been
prepared with a velocity selection pulse of length $\tau_s$, as
discussed in Sec.~\ref{ss:hoc}, on the average
$|\delta(t')| \sim |\delta(0)| \sim 1/\tau_s$ so we need again to
compare ${\cal V}^\pm(t)$ with $\delta{\cal V}^\pm(t)$ where the
increment is on a distance of the order $v_r\tau_s$. For our typical
numbers such increment is of the order of 1 $\mu$m and, as above, the
considerations on the smoothness of ${\cal V}^\pm$ over a short distance
can be applied.

\section{\label{app:wig}Equivalence with the Wigner function formalism}
A review on the Wigner functions and quantum mechanics in phase space
can be found in \cite{Curtright_2014}, and its specific application in
atom interferometry in \cite{Giese_2015}. Here we briefly summarize
and compare previous results to ours. To avoid confusion, we restore
hats to distinguish the operators from the variables of the Wigner function.

We point out that a convenient starting point for evaluating the
Wigner function is not the initial Hamiltonian in Eq.~(\ref{eq:h0})
but rather Eq.~(\ref{eq:h3}), with the time dependent operators
$\hat{Z} = \hat{z}(t)$ and $\hat{P}=\hat{p}(t)-m\Delta(t)/k$, with
$\hat{z}(t),\hat{p}(t)$ defined in Eq.~(\ref{eq:zt}) and
Eq.~(\ref{eq:pt}) respectively. The two operators obey the canonical
commutation relation $[\hat{Z},\hat{P}]=i\hbar$, so we can use the
Weyl-transforms of $\hat{Z},\hat{P}$ as coordinates $z,p$ in phase
space. Moreover, since the transformation
$\hat z\rightarrow \hat Z, \hat p\rightarrow \hat P$ is linear,
it is not only mapping the Heisenberg equation onto the Moyal equation
but also acts as a coordinate change in phase space \cite{Dias_2004}.

Here we show that neglecting $\hat{H}_S^{\rm{III}}$ in
Eq.~(\ref{eq:h3}) leads readily to Eq.~(\ref{eq:prob21bis}) and
Eq.~(\ref{eq:epsV}) also in phase space.

To simplify the notation we introduce the spinorial Wigner functions
associated to a generic initial density matrix, $\hat\rho(0)$:
\begin{equation}
  W_{jk}(z,p;0)\equiv \displaystyle{\frac{1}{2\pi}\int
    e^{-ipu}\bra{z-\frac{\hbar}{2}u}\hat\rho_{jk}(0)\ket{z+\frac{\hbar}{2}u}\,du},
  \label{eq:fdef}
\end{equation}
where the indices $j,k$ refer to the spinorial component in the basis
$\ket{\pm}$ defined in Eq.~(\ref{eq:eigen3}).

In our case, at $t=0$ the spatial wavefunction is $\psi(z)$ and the spinorial state is $\ket{1}=(\ket{+}+\ket{-})/\sqrt{2}$, thus the density matrix 
$\hat \rho(0)=\ket{\psi;1}\bra{\psi;1} $ corresponds to a $2\times2$
Wigner function 

\begin{equation}
W_{jk}(z,p;0)=\frac{1}{2}f(z,p) 
\end{equation}
with 
\begin{equation}
f(z,p)= \displaystyle{\frac{1}{2\pi}\int
    e^{-ipu}\psi^*\left(z-\frac{\hbar}{2}u\right)\psi\left(z+\frac{\hbar}{2}u\right)\,du}.
\end{equation}
Note that $f(z,p)$ is real.

The temporal evolution of $W$ obeys the Moyal equations
\cite{Curtright_2014}
\begin{equation}
  \partial_t W_{jk}=\frac{h_{jj} \star W_{jk}-W_{jk} \star h_{kk}}{i \hbar}
  \label{eq:Meq}
\end{equation}
where the $\star$-product is defined as
\begin{equation}
  a \star b=a\, \exp \left[\frac{i \hbar}{2}
        \left(
      {\overleftarrow \partial_z}{\overrightarrow\partial_p}-
      {\overleftarrow\partial_p}{\overrightarrow\partial_z}
    \right)
  \right]\,b,
\end{equation}
with the arrows indicating if the derivative operators act on $a$ or
$b$, and $h(t)=p(t) v_r\sin \phi_1(t)\sigma_3/2$ is the Weyl-transform
of the Hamiltonian $\hat H=\hat{P} v_r\sin \phi_1(t)\sigma_3/2$ in the
$\ket{\pm}$ basis.

When $h$ is at most quadratic in $z$ and $p$ -- actually just
proportional to $p$ in our case -- Eq.~(\ref{eq:Meq}) involving
the diagonal elements of $W$ simplifies to the Liouville equation.
The solutions for $W_{jj}$ are then
$W_{jj}(z,p,t) = f(z-z_j^c,p-p^c)$ where $z_j^c(t),p^c(t)$
indicate the classical trajectories, namely
\begin{equation}
  \begin{cases}
    z_\pm^c(t)=\pm\displaystyle{\frac{v_r S(t)}{2}}\\
    \\
    p^c(t)=\displaystyle{p^c(0)=\frac{m\Delta(0)}{k}}.
  \end{cases}
\end{equation}
For the off-diagonal elements we have
\begin{equation}
  W_{jk}(z,p;t)=\exp[i\alpha_k(t)-i\alpha_j(t)]f(z,p),
\end{equation}
with
\begin{equation}
  \alpha_\pm(t)=\frac{1}{\hbar}\int_0^t E^\pm(t')\,dt',
  \label{eq:alphaj}
\end{equation}
and $E^\pm$ from Eq.~(\ref{eq:pmeigenvals}).

With $W(z,p;2T)$ we can calculate the transition probability
\begin{equation}
\begin{split}
P_{21}&=\mathrm{Tr}\left[\hat \rho(2T)\ket{2}\bra{2}\,\right] \\
&= \frac{1}{2} \int \mathrm{Tr}\left[ W(z,p;2T)
\begin{pmatrix}
\phantom{-}1 & -1 \\
-1 & \phantom{-}1
\end{pmatrix}\right] \,dz\,dp \\
&=\frac{1}{2}\int \left[ W_{\!_{++}} + W_{\!_{--}} - 2\mathrm{Re}(W_{\!_{+-}}) \right]\,dz\,dp\\
&=\frac{1-\cos \Delta \alpha(2T) }{2},
\end{split}
\end{equation}
with $\Delta \alpha=\alpha_+-\alpha_-$, which is the same as
Eq.~(\ref{eq:prob21bis}).

When the weak potential ${\cal V}$ is added to $h(t)$ in Eq.~(\ref{eq:h3p}),
the correction $\Delta^{(1)} \alpha(t)$ in Eq.~(\ref{eq:alphaj}) is
obtained by applying perturbation theory at the first order as outlined in \cite{Curtright_2014},\cite{Curtright_2001}
\begin{equation}
  \begin{split}
    \Delta^{(1)}\alpha(t)=&\frac{1}{\hbar}\int {\cal V}(z)[W_{\!_{++}}(z,p,t)-W_{\!_{--}}(z,p,t)]
    \,dz\,dp\\
    \simeq & \frac{{\cal V}^-(z,z_{\!_+}^c)}{\hbar}
  \end{split}
\end{equation}
with ${\cal V}^-(z,\Delta z)$ defined as in Eq.~(\ref{eq:vpm}), where
we assumed that $W_{jj}$ is localized around $z_j^c$ and acts as $\delta(z-z_j^c)$ in the integral. 
The result is a correction
term to $P_{21}$ which is the same as $2\epsilon_2$ in
Eq.~(\ref{eq:epsV}).

The corrections to the classical trajectories $z_\pm^c(t)$ due to
$\cal V$ can instead be neglected at the first order in $\cal V$
according to the general rules of variational calculus, as discussed
in \cite{Greenberg_2012}.

%

\end{document}